\documentstyle[aps]{revtex}
\begin{document}
% \draft command makes pacs numbers print
\draft
\preprint{}

\title{The Application of the Newman-Janis Algorithm in
Obtaining Interior Solutions of the Kerr Metric.}
\author{S. P. Drake\thanks{Permanent address:
Department of Physics and Mathematical Physics,
University of Adelaide, Adelaide S.A. 5005, Australia.}
\thanks{email: sdrake@phyiscs.adelaide.edu.au}
$^{1,2}$ and R. Turolla$^1$}
\address{ $^1$ Department of Physics, University of Padova, Via 
Marzolo 8, 35131 Padova, Italy.}  
\address{ $^2$ School of Physics, University of Melbourne, Parkville
Victoria 3052, Australia.}
\date{\today}
\maketitle
\begin{abstract}
In this paper we present a class of metrics to be considered as 
new possible sources for the Kerr metric. These new solutions are 
generated 
by applying 
the Newman-Janis algorithm (NJA) to any static spherically symmetric 
(SSS) ``seed'' metric.
The continuity conditions for joining any two of these new metrics 
is presented. A specific analysis of the joining of interior 
solutions to the  
Kerr exterior is made. 
The boundary conditions used are those first developed by 
Dormois and Israel. 
We find that the NJA can be used to generate new physically 
allowable 
interior solutions. These new solutions can be matched smoothly 
to the Kerr metric. 
We present a general method for 
finding such solutions with oblate spheroidal boundary surfaces. 
Finally a trial solution is found and presented. 
\end{abstract}
\section{Introduction} \label{sec:i}
Since the discovery of the Kerr metric~\cite{pap:kerr} many attempts 
have been made to find a physically reasonable interior matter 
distribution
that may be considered as its source. For a review of some of these 
approaches the reader is referred to the introductions 
of~\cite{pap:dmm}
and~\cite{pap:ak}. Though much progress has been made results 
have been generally disappointing.  
As far as we can tell nobody  has obtained a physically satisfactory 
interior solution. This seems surprising given the 
success of matching internal spherical symmetric 
solutions to the Schwarzschild metric. The problem is not 
simply that the loss of one degree of symmetry makes 
the derivation of analytic results that much more difficult. Severe
 restrictions
are placed on an interior metric by maintaining that it 
must be joined smoothly to the Kerr metric. Further restrictions
are placed on interior solutions to ensure that they correspond to 
 physical objects. Furthermore
since the Kerr metric 
has no radiation 
field associated with it its source metric must also be 
non-radiating. This places even further constraints on the structure
of the interior metric~\cite{bk:dk}. 
Given the strenuous nature of these limiting conditions is not 
surprising to learn that as yet nobody obtained 
a truly satisfactory solution to the 
problem of finding  sources for the Kerr metric. 
In general the failure is  due 
to an internal structure that has unphysical properties, or 
a failure to satisfactorily match the boundary conditions.  

This paper tries to overcome some these problems by examining a 
broad class 
of interior metrics which, by construction, are matched 
smoothly to the 
Kerr metric. Concurrently all the sources that we will be examining 
are stationary and 
axially symmetric (SAS) and as a consequence they are not generators 
of gravitational radiation.
The boundary conditions used, those first 
discussed by Dormois~\cite{pap:gd} and developed further  
by Israel~\cite{pap:is}, are explained in more 
detail later. At this stage we simply point out that in order to 
evaluate the Dormois-Israel conditions a knowledge
of the surface separating the two space-times (interior and exterior)
is required. Unlike many previous attempts, however, we do not 
specify the 
surface prior to evaluating the Dormois-Israel conditions. 
In fact we take the reverse approach, that is 
the surface between the 
two metrics is determined by imposing the Dormois-Israel criteria. 
This then allows us to explore all possible surfaces on which 
the two metrics can be matched continuously.  
To the best of the authors knowledge this
approach has not been previously examined.  

After the original discovery of the Kerr metric
Newman and Janis showed that this result could 
be ``derived'' by making an elementary complex 
transformation to the Schwarzschild solution~\cite{pap:nj1}.  
This same method was then used to obtain a new  
SAS solution to Einstein's field equations
now known as the Kerr-Newman metric~\cite{pap:nj2}. 
The Kerr-Newman space-time is that associated with the exterior 
geometry of a rotating massive charged black-hole. For a modern review 
of the Newman-Janis algorithm (NJA)  
for obtaining both the Kerr and Kerr-Newman metrics see~\cite{bk:ier}.

At the time of publication 
there was no valid reason as to why this method worked.
Many physicist considered this {\it ad hoc} procedure to 
be a ``fluke'' and not worthy of further investigation. 
However by means of a very elegant mathematical 
method, Schiffer et al~\cite{pap:mms} gave a rigorous proof as to 
why the Kerr metric can be considered as a complex transformation
of the Schwarzschild space-time. We will not go into 
the details of this paper, but simply state that it tells us 
nothing about the success of such a complex transformation on a spherically 
symmetric interior solution to give an axially symmetric one. 
The reason for this is that the proof they gave 
relies on two assumptions. The first assumption is that the metric
belongs to the same algebraic class as the Kerr-Newman solution, namely 
the Kerr-Schild class (KS)~\cite{pap:gcd}. The second
is that the metric corresponds to a empty solution of Einstein's field 
equations. In the case we study these assumptions are 
not made and hence the proof is not applicable. It is clear, by the 
generation of the Kerr-Newman metric, that all the components of the
stress-energy tensor need not be zero for the NJA to be successful. 
In fact G\"{u}rses and G\"{u}rsey in 1975~\cite{pap:mg} 
showed that if a metric can be written in KS form then a complex
transformation ``is allowed in general relativity.'' 

Since it is not possible to represent a perfect fluid
(with the exception of the cosmological fluid) by a KS 
space-time~\cite{pap:gm2},
there is no proof as to the validity of complex transformations 
in generating physically reasonable stationary axially symmetric (SAS) 
interior 
solutions from spherically symmetric ones. This, of course, does 
not imply that the complex transformation 
 method is valid only on KS space-times.
It simply means we have no conception as to whether or not 
it will yield sensible results. Recently KS geometries have been 
considered in 
looking for elastic-solid
bodies as sources for the Kerr metric~\cite{pap:gm1}~\cite{pap:gm2}. 
This approach seems promising but preliminary inquiries
show that the sources are either surrounded
 by a shell-like distribution of 
stress energy or else exhibit a ring singularity. For these 
reasons we feel 
that it is important to investigate the nature of complex coordinate 
transformations in non KS space-times. 

There are two main themes that run through this paper. The first 
is the 
application of the NJA to any static spherically symmetric (SSS) 
seed metric. 
The second is the joining of two SAS metrics 
on a static axially symmetric boundary surface. These two themes are 
linked together with the aim of finding a physically reasonable source 
for the Kerr metric. As far as the authors are aware this combined 
approach is entirely
new.

The rest of the outline of the paper is as follows. In the second 
section~\ref{sec:tnja} we briefly review the Newman-Janis algorithm 
NJA for obtaining SAS metrics from SSS ones. The 
resulting metric is written in 
terms of two arbitrary 
functions. These functions give the physical properties of the 
internal structure. Furthermore 
we make use of a coordinate transformation so that this 
new metric is written in Boyer-Lindquist type coordinates. This makes 
the physical interpretation much clearer and decreases the 
amount of algebra required in calculating various metric 
properties. The third section~\ref{sec:tmotstoacbs}
 outlines 
the boundary conditions used in the Dormois-Israel formalism. In 
this section we develop a set of boundary conditions for the joining 
of any two SAS metrics on a arbitrary static axially 
symmetric boundary surface. The term boundary surface is used 
according to Israel's original definition. This means that 
the surface separating the two geometries has a vanishing 
surface stress-energy tensor, i.e. no thin shells. 
In this section we place particular emphasis on those 
SAS metrics that can be generated from SSS metrics via 
the NJA. Following on directly from this section~\ref{sec:psftkm} 
examines explicitly the case when the exterior metric 
corresponds to the Kerr metric. In this section the boundary 
conditions for matching of an interior generated by the 
NJA to the Kerr metric are given. Also in this section we discuss 
in what sense we use the term ``physically reasonable source of the 
Kerr metric''. This places even further constraints on the interior 
metric form. Having a established a rigorous formalism 
in the previous 
sections in the next section~\ref{sec:ats} we 
look for solutions to which
the NJA may be successfully applied. We examine one such
solution which we have called the ``trial solution''. 
We find that it has 
a physically reasonable seed metric associated with it and 
matches smoothly to the Kerr metric on an oblate spheroid. 
The last section~\ref{sec:c} is the conclusion and sums up all the 
new results that have been obtained in this paper. 
 
\section{The Newman-Janis Algorithm.} \label{sec:tnja}

We do not claim to be the first authors to examine SAS 
interior solutions generated by the NJA. In fact a  paper by 
Herrera and 
Jim\'{e}nez~\cite{pap:lh} obtains the same general metric form after 
the NJA is applied as we do. However the similarity ends there.
In our results 
the metric is cast in a more tangible form, we use a different set of
boundary conditions and we examine different specific metric 
functions. 

Despite the work by Newman and Janis, the work by Herrera and 
Jim\'{e}nez and some further papers on the subject the NJA is not 
a well known area of general relativity. For this reason 
we believe it is necessary to give a detailed outline 
as to what it actually is. 
We start by reviewing this algorithm as applied to a static 
spherically symmetric ``seed'' metric,
\begin{equation}
ds^2 = e^{2\phi(r)}dt^2 - e^{2\lambda(r)}dr^2 - 
r^2(d\theta^2 + \sin^2\theta d\Phi^2), \label{eqn:ssm}
\end{equation}
with the signature chosen to be consistent with Newman and Janis' 
original paper.
Following Newman and Janis, equation~(\ref{eqn:ssm}) is written 
in advanced 
Eddington--Finkelstein coordinates, i.e. 
the $g_{rr}$ component is eliminated by a change of coordinates and 
a cross term is introduced. 
Specifically this is done by advancing the time 
coordinate so that $dt = du + fdr$
and setting $f = \pm e^{\lambda(r)-\phi(r)}$, we choose the 
positive case, again to be consistent with the 
Newman-Janis formulation. Once this is done the metric in these new 
coordinates is,
\begin{equation}
ds^2 = e^{2\phi(r)}du^2 + 2e^{\lambda(r)+\phi(r)}dudr - 
r^2(d\theta^2 + \sin^2\theta d\Phi^2).
\end{equation}

Written in contravariant form this is
\begin{equation}
g^{\mu\nu} = \left(
\begin{array}{cccc}
0 & e^{-\lambda(r)-\phi(r)} & 0 & 0 	\\
e^{-\lambda(r)-\phi(r)} & -e^{-2\lambda(r)} & 0 & 0 \\
0 & 0 & -1/r^2 & 0 \\
0 & 0 & 0 & -1/(r^2\sin^2{\theta})
\end{array}    \right).
\end{equation}
This is done so that the above metric may be written in 
the terms of its 
null tetrad vectors, 
\begin{equation}
g^{\mu\nu} = l^\mu n^\nu + l^\nu n^\mu-m^\mu\bar{m}^\nu 
- m^\nu\bar{m}^\mu, \label{eq:gmet}
\end{equation}
where
\begin{eqnarray}
l^\mu  & = & \delta^\mu_1 	\\
n^\mu & = & -\frac12 e^{-2\lambda(r)}\delta^\mu_1 
+ e^{-\lambda(r)-\phi(r)}\delta^\mu_0 \\
m^\mu & = & \frac{1}{\sqrt{2}\bar{r}}
\left( \delta^\mu_2 
+ \frac{i}{\sin{\theta}}\delta^\mu_3 \right).
\end{eqnarray}
The bar indicates a complex conjugate. This complex null tetrad 
system forms the starting point for the ``derivation''
of Kerr-Newman space-times. As has been already stated 
this procedure is known to be a valid method for KS
geometries but its extension into non KS type metrics is still to 
be thoroughly examined.  
To be consistent exactly the same transformations 
as those originally performed by Newman and Janis are made. That is
coordinates are advanced by the following complex increments:
\begin{equation}
u  \rightarrow  u' = u - ia\cos{\theta}, \;
r  \rightarrow  r' = r + ia\cos{\theta}, \; 
\theta \rightarrow \theta', \; \Phi \rightarrow \Phi'. \label{eq:trans}
\end{equation}
By keeping $r'$ and $u'$ real (that is considering the 
transformations as a complex rotation of the 
$\theta,\Phi$ planes) one
obtains the following tetrad.  
\begin{eqnarray}
l^\mu & = & \delta^\mu_1 \\
n^\mu & = & -\frac12 e^{-2\lambda(r,\theta)}\delta^\mu_1 +
 e^{-\lambda(r,\theta) - \phi(r,\theta)}\delta^\mu_0 \\
m^\mu & = & \frac1{\sqrt{2}(r + ia\cos{\theta})}\left(
ia\sin{\theta}(\delta^\mu_0 - \delta^\mu_1) + 
\delta^\mu_2 + \frac{i}{\sin{\theta}}\delta^\mu_3\right)
\end{eqnarray}
All primes are now dropped for convenience
of notation but one must recall that the new 
functions $e^{\lambda(r,\theta)}$ 
and $e^{\phi(r,\theta)}$ are not the same as the old ones. 
In fact, the new 
functions depend on both $r$ and $\theta$ whereas the 
old ones had only an $r$ dependence. 

The metric formed from the above null vectors 
using~(\ref{eq:gmet}) is,
\begin{equation}
g^{\mu\nu} = \left(
\begin{array}{cccc} \label{eqn:newmetric}
-\frac{a^2 \sin^2{\theta}}{\Sigma} & e^{-\lambda(r,\theta)-\phi(r,\theta)} + 
\frac{a^2\sin^2{\theta}}{\Sigma} & 0 & -\frac a\Sigma \\
. & - e^{-2\lambda(r,\theta)} - \frac{a^2\sin^2{\theta}}{\Sigma} 
& 0 & \frac{a}{\Sigma} \\
. & . & -\frac{1}{\Sigma} & 0 \\
. & . & . & -\frac{1}{\Sigma\sin^2{\theta}} \\
\end{array}      \right)
\end{equation}
where $\Sigma = r^2 + a^2\cos^2{\theta}$.
In the covariant form this is 
\begin{equation} \label{eqn:coform}
g_{\mu\nu} =      \left(
\begin{array}{cccc}
e^{2\phi(r,\theta)} & e^{\lambda(r,\theta) + \phi(r,\theta)} & 0 & a\sin^2{\theta}
e^{\phi(r,\theta)}(e^{\lambda(r,\theta)}- e^{\phi(r,\theta)}) \\
. & 0 & 0 & -ae^{\phi(r,\theta)+\lambda(r,\theta)}\sin^2{\theta} \\
. & . & -\Sigma & 0 \\
. & . & . & -\sin^2{\theta}(\Sigma + 
a^2\sin^2{\theta}e^{\phi(r,\theta)}(2e^{\lambda(r,\theta)}-
e^{\phi(r,\theta)})) \\    
\end{array}      \right)
\end{equation}

As the metric is symmetric the ``.'' is used to indicate $g^{\mu\nu} 
= g^{\nu\mu}$. The form of this metric gives the general result of the NJA 
to {\it any} SSS seed metric.    

The metric given in equation~(\ref{eqn:coform}), though relatively 
simple, is still hard to work with. To eradicate this problem 
one can make a gauge transformation  
so that the only off-diagonal component is $g_{\Phi t}$.
This makes it easier to compare with the more usual Boyer-Lindquist 
form of the Kerr metric~\cite{pap:boy} and to interpret physical 
properties such as frame dragging. 
It also aids in the calculation and evaluation of the Einstein tensor.
To do this, the coordinates $u$ and $\Phi$ are redefined in 
such a way that the metric in the 
new coordinate system has the properties 
described above. More explicitly, if we advance the 
coordinates in the following way, 
$du  =  dt + g(r)dr $ and
$d\Phi = d\Phi' + h(r)dr$,   
with the following functional forms of $f$ and $g$:
\begin{eqnarray}
g(r) & = & -\frac{e^{\lambda(r,\theta)}
(\Sigma + a^2\sin^2{\theta}e^{\lambda(r,\theta) + \phi(r,\theta)})}
{e^{\phi(r,\theta)}(\Sigma + a^2\sin^2{\theta}e^{2\lambda(r,\theta)})} \\
h(r) & = & -\frac{ae^{2\lambda(r,\theta)}}
{\Sigma + a^2\sin^2{\theta}e^{2\lambda(r,\theta)}},
\end{eqnarray}
then after some algebraic manipulations one finds that in 
this coordinate system the metric is,
\begin{equation} \label{eq:blform}
g_{\mu\nu} =      \left(
\begin{array}{cccc}
e^{2\phi(r,\theta)} & 0 & 0 & a\sin^2{\theta}
e^{\phi(r,\theta)}(e^{\lambda(r,\theta)}- e^{\phi(r,\theta)}) \\
. & -\Sigma/(\Sigma e^{-2\lambda(r,\theta)} + a^2\sin^2{\theta})
 & 0 & 0 \\
. & . & -\Sigma & 0 \\
. & . & . & -\sin^2{\theta}(\Sigma + 
a^2\sin^2{\theta}e^{\phi(r,\theta)}(2e^{\lambda(r,\theta)}-
e^{\phi(r,\theta)})) \\    
\end{array}      \right). \label{eq:boy}
\end{equation}

We state that this metric represents the complete family of
metrics that may be obtained by performing the NJA on any
static spherical symmetric seed metric, written in Boyer-Lindquist
type coordinates. The validity of these transformations requires
that $\Sigma + a^2\sin^2\theta e^{2\lambda(r,\theta)} \neq 0$,
which is always the case since $e^{2\lambda(r,\theta)}> 0$.
We note here that the choice of $e^{2\phi(r,\theta)}= 
e^{-2\lambda(r,\theta)} = 1 + (Q^2 -2mr)/\Sigma$ 
corresponds to the Kerr-Newman solution, where $Q$ and $m$ are 
the charge and mass of the body respectively.

\section{The matching of two space-times 
on a common boundary surface}
\label{sec:tmotstoacbs}
The problem of matching two separate space-times on
a common surface is a well explored area of general relativity. 
A large number of papers exist on the subject, and now  
an algebraic programme~\cite{prog:pm} is available to speed 
up calculations.
However Israel's original paper~\cite{pap:is} still remains a 
definitive work on the subject. Although the 
formalism was originally developed to examine the motion 
of expanding bubbles in the universe it has 
applications that range far wider.  
We will not go into a detailed discussion of the Dormois-Israel 
formalism, 
we will simply use it as it was 
originally developed. 

When matching to two different metrics on a common surface the main problem 
one faces is that of choosing an appropriate coordinate system. One needs to be convinced that
if any discontinuities do occur they are due solely to the topology and 
not the coordinate systems. To eliminate 
this problem a common coordinate system must be
defined on the surface. It is argued that in this new 
coordinate system the 
continuous properties should be the metric coefficients along with the 
extrinsic curvature (or second fundamental form) as evaluated at the surface. 

In our case an attempt is made to join an interior 
solution, which acts as a source, to the Kerr metric. It seems reasonable 
then to assume that the surface dividing the two space-times is both static 
and axially symmetric. With this in
mind we examine how the Dormois-Israel formalism can be simplified and
what restrictions it places on both the surface and the 
internal metric. 

The approach taken here is as follows; one starts by specifying
the coordinates on which the surface is defined.
Let us suppose that both the interior and exterior metrics are 
defined in the coordinate system $x^\alpha = \{t,r,\theta,\Phi\}$.
Let us further suppose that the boundary surface separating 
these two space-times 
is a function connecting the two variables
$r$ and $\theta$. This simply states that the separating surface is
 both 
static and axially symmetric. If this is true then the two coordinates 
$r$ and $\theta$ may be eliminated by a single parameter $\tau$. 
As a consequence of this, the surface coordinates 
may be expressed as  $\zeta^i = \{t,\tau,\Phi\}$. These are the 
types of
boundary surfaces that we study. 
If in general, by the definition of a boundary surface, two 
coordinates can be 
replaced by a single one at the surface, then the four metric
$^4g_{\alpha\beta}$ may be replaced by the three metric $^3g_{ij}$.
Note the we are using the convention that Greek indices run 
from 0 to 3 and Latin 
from 0 to 2. The four metric is then projected onto the surface 
three metric by the following relationship,
\begin{equation}
^3g_{ij} = \frac{\partial x^\alpha}{\partial \zeta^i} \label{eq:met}
\frac{\partial x^\beta}{\partial \zeta^j}\,^4g_{\alpha\beta}.
\end{equation}

It is handy at this stage to introduce the following notation;
let $M^+$ correspond to the space-time geometry exterior to the 
object we are examining. In the case we are most interested in,
$M^+$ corresponds to Kerr 
geometry. The interior geometry is given the 
symbol $M^-$. For any metric-dependent quantity $X$ one must specify 
the region of space-time in which it is to be calculated. The notation 
$X^+$ means that the quantity $X$ is calculated in the exterior 
space-time geometry $M^+$. The notation $X^+\!\mid_s$ signifies that the 
value $X$ is calculated in $M^+$ and evaluated at the 
surface. Finally we use the notation that 
$[X] \equiv X^+\!\mid_s - X^-\!\mid_s$, which measures the jump 
discontinuity in the value of $X$ as calculated by the two metrics 
and evaluated at the surface.   

Israel's first boundary condition demands the continuity of the 
first fundamental form, that is $[^3g_{ij}] = 0$.
If the two metrics to be joined may both be written in a form in 
which the only off-diagonal term is $g_{t\Phi}$ then 
the above results 
reduce to 
\begin{eqnarray}
\left[ ^3g_{tt}\right] & =  & \left[^4g_{tt}\right] = 0 \label{eq:beg} \\
\left[^3g_{t\Phi}\right] & = &  \left[^4g_{t\Phi}\right]= 0\\
\left[^3g_{\Phi\Phi}\right] & =  & \left[^4g_{\Phi\Phi}\right] = 0\\
\left[^3g_{\tau\tau}\right] & = & \left[\left(\frac{\partial r}{\partial\tau}
\right)^2\,^4g_{rr} + \left(\frac{\partial \theta}{\partial\tau}
\right)^{2}\,^4g_{\theta\theta}\right]= 0.  \label{eq:end}
\end{eqnarray}
All these four equations when used in conjunction 
with the metric~(\ref{eq:boy}) are simplified to the 
two conditions
\begin{equation}
\left[e^{2\lambda(r,\theta)}\right] = 
\left[e^{2\phi(r,\theta)}\right]  =  0. \label{eq:con1}
\end{equation}
 The above equations give the continuity of metric coefficients at the 
boundary surface. 

The next of Israel's boundary conditions involves the 
properties of the extrinsic curvature. Most work done with 
the matching of extrinsic curvature has been based on the thin
shell formalism (for examples see~\cite{bk:MTW} to \cite{pap:ll}), however 
the general technique may be used in a much wider range of 
scenarios. The primary 
result that we are interested in is that for boundary surfaces
(i.e. surfaces that have a surface energy-momentum tensor which is 
identically zero) all components of the 
extrinsic curvature are continuous at the surface.
 
The extrinsic curvature measures the rate of change of the 
 normal vector as it moves along the boundary surface. It is 
given explicitly by the expression 
\begin{eqnarray}
K_{ij} &  = & n_{i;j}  \\
& = & -n_\gamma \left( \frac{\partial^2 x^\gamma}{\partial\zeta^i
\partial\zeta^j} + \Gamma^\gamma_{\alpha\beta}
\frac{\partial x^\alpha}{\partial\zeta^i}\frac{\partial x^\beta}
{\partial\zeta^j}\right), \label{eq:ex}
\end{eqnarray}
where $n^\gamma$ is the unit normal to the surface and 
$\Gamma^\gamma_{\alpha\beta}$ are the
Christoffel symbols associated with a given metric. The components
of the normal $n^\gamma$ clearly depend on the hyper-surface 
separating the two space-time regions $M^+$ and $M^-$. In general 
if the surface is specified by the equation
\begin{equation}
F(x^\alpha) = 0 
\end{equation}
then the components of the normal vector are
\begin{equation}
n_\gamma = \pm\frac{\partial_\gamma F}{\sqrt{\partial_\beta 
F \partial^\beta F}}.
\end{equation}
The $\pm$ determines whether the normal is a space-like
or time-like vector.
In the case of static axially symmetric surfaces then 
\begin{equation}
F = r - R(\theta) = 0. \label{eq:bound}
\end{equation}
$R(\theta)$ is some unknown function of $\theta$ which specifies 
the boundary surface. In this case the unit normal, 
which we consider to be a time-like vector, is given by 
\begin{equation}
n_\gamma = - \frac{\delta^1_\gamma - \partial_\theta R(\theta)\delta^2_\gamma}
{\sqrt{g^{rr} - g^{\theta\theta}(\partial_\theta R(\theta))^2}}.\label{eq:norm}
\end{equation}

Using the above equation along with~(\ref{eq:ex}) and~(\ref{eq:beg})
to~(\ref{eq:end}) one finds that the matching of extrinsic curvature 
reduces to the following set of constraints on the metric
components
\begin{eqnarray}
\left[K_{tt}\right] & = &  g^{rr}\left[g_{tt,r}\right] + 
g^{\theta\theta} \partial_\theta R(\theta)
\left[g_{tt,\theta}\right] = 0   \label{eq:con2} \\
\left[K_{t\Phi}\right] & = & g^{rr}\left[g_{t\Phi,r}\right] + 
g^{\theta\theta} \partial_\theta R(\theta)
\left[g_{t\Phi,\theta}\right] = 0 \\  \label{eq:con3}
\left[K_{\Phi\Phi}\right] & = & g^{rr}\left[g_{\Phi\Phi,r}\right] + 
g^{\theta\theta} \partial_\theta R(\theta)
\left[g_{\Phi\Phi,\theta}\right] = 0  \label{eq:con4} \\
\left[K_{\tau\tau}\right] & = & \frac12\left(\frac{\partial r}
{\partial \tau}\right)^2
\left( g^{rr}\left[g_{rr,r}\right]  + \partial_\theta R(\theta)
g^{\theta\theta}\left[g_{rr,\theta}\right]\right)  \nonumber \\
& + & \frac{\partial r}{\partial \tau}\frac{\partial 
\theta}{\partial \tau}
\left( g^{rr}\left[g_{rr,\theta}\right]  - \partial_\theta R(\theta) 
g^{\theta\theta}\left[g_{\theta\theta,r}\right]\right) 
\label{eq:con5} \\
& - & \frac12 \left( \frac{\partial \theta}{\partial \tau}\right)^2
\left( g^{rr}\left[g_{\theta\theta,r}\right]  
+ \frac{\partial R(\theta)}{\partial \theta}
g^{\theta\theta}\left[g_{\theta\theta,\theta}\right]\right)
=0.\nonumber
\end{eqnarray}
In the above equations all metric components refer to the 
four metrics 
$^4g_{\alpha\beta}$ and are evaluated at the surface.
The notation $X,_\alpha \equiv \partial_\alpha X \equiv 
\frac{\partial X}{\partial x^\alpha}$ is used. 
Note also that since all components of the 
extrinsic curvature are zero common factors have 
been removed. 
If the two metrics to be joined can be generated by the 
NJA and hence may me written in the form~(\ref{eq:boy}) 
then equations~(\ref{eq:con2}) to~(\ref{eq:con4}) may be greatly simplified. 
Assuming $a \neq 0$ we find that above equations are satisfied by 
\begin{eqnarray}
g^{rr}\left[e^{2\phi(r,\theta)},_r\right] & - &
\partial_\theta R(\theta) g^{\theta\theta}\left[
e^{2\phi(r,\theta)},_\theta\right] = 0 \label{eq:sim1} \\
g^{rr}\left[e^{2\lambda(r,\theta)},_r\right] & - &
\partial_\theta R(\theta) g^{\theta\theta}\left[
e^{2\lambda(r,\theta)},_\theta\right] = 0. \label{eq:sim2}
\end{eqnarray}
The common factors mentioned above involve the 
rotation parameter $a$. When this is zero the only surviving 
component of the extrinsic curvature is $K_{tt}$. If this is the
case the boundary conditions are expressed 
simply by~(\ref{eq:sim1}) as expected. 
If $\partial_\theta R(\theta)\neq0$  
then equation~(\ref{eq:con5}) is simplified  
using the relations~(\ref{eq:boy}) and~(\ref{eq:bound}) to
read
\begin{equation}
\partial_\theta R(\theta)
\left( g^{rr}\left[g_{rr,r}\right]  + 
g^{\theta\theta}\left[g_{rr,\theta}\right]\right)
+ 2g^{rr}\left[g_{rr,\theta}\right] =0
\label{eq:sim3}.
\end{equation} 
However if  $\partial_\theta R(\theta)=0$ 
then $[K_{\tau\tau}\equiv0]$.

The above 
equations~(\ref{eq:con1}) and~(\ref{eq:sim1}) to 
~(\ref{eq:sim2}) form a complete set 
of boundary conditions for the joining of any two stationary axially 
symmetric metrics generated by the 
NJA  when applied to any SSS seed metric on 
an axially symmetric boundary surface.
%By the term ``boundary surface'' it is meant that the 
%surface separating 
%the two space-times geometries has a surface energy density 
%tensor which is 
%identically zero. 
In the next section we 
examine some of the properties of these continuity conditions
when the exterior metric is that due to Kerr.

\section{Possible sources for the Kerr metric}
\label{sec:psftkm}
The ultimate goal of this paper is find new static axially symmetric 
solutions to Einstein's field equations which may be considered 
as sources for the Kerr metric. With this 
in mind we need to examine the 
properties of ~(\ref{eq:con1}),~(\ref{eq:sim1}) 
and~(\ref{eq:sim2}) when the exterior metric is Kerr. 
In this case the boundary conditions become,
\begin{eqnarray}
e^{2\phi(r,\theta)-}\!|_s & = & 1 - \frac{2MR(\theta)}{\Sigma_s}, 
\label{eq:k1}  \\
e^{2\lambda(r,\theta)-}\!|_s & = & \frac{\Sigma_s}{\Sigma_s - 2MR(\theta)},
\label{eq:k2}  \\
\frac{2M\left( R(\theta)^2-a^2\cos^2{\theta}\right) }{\Sigma^2_s} & - &
e^{2\phi(r,\theta)-},_r\!|_s  =  \nonumber  \\
 - \frac{\partial_\theta R(\theta)}
{\Delta_s}\left( \frac{4a^2MR(\theta)\cos{\theta}\sin{\theta}}{\Sigma^2_s}
\right. & + &
\left. e^{2\phi(r,\theta)-},_\theta\!|_s \right),  \label{eq:k3} \\   
\frac{2M(a^2\cos^2\theta -R^2(\theta))}{\Delta^2_s} & - & 
e^{2\lambda(r,\theta)-},_r\!|_s   =  \nonumber \\
 \frac{\partial_\theta R(\theta)}
{\Delta_s} \left( \frac{4a^2MR(\theta)\cos{\theta}\sin{\theta}}{\Delta^2_s} 
\right. & - & 
\left. e^{2\lambda(r,\theta)-},_\theta\!|_s \right),  \label{eq:k4}\\
\partial_\theta R(\theta)\left(\frac{2M\left( R(\theta)^2-a^2
\cos^2{\theta}\right) }{\Sigma^2_s}\right.  & - &\left.
 e^{-2\lambda(r,\theta)-},_r\!|_s  \right) =  \nonumber  \\
 - \frac{\partial_\theta R(\theta) + 2}{\Delta_s}
\left( \frac{4a^2MR(\theta)\cos{\theta}\sin{\theta}}{\Sigma^2_s}
\right. & + &
\left. e^{-2\lambda(r,\theta)-},_\theta\!|_s \right),  
\label{eq:k5}
\end{eqnarray}
where
\begin{eqnarray}
\Delta_s & = & R(\theta)^2 - 2MR(\theta) + a^2 \\
\Sigma_s & = & R(\theta)^2 + a^2\cos^2{\theta}.       
\end{eqnarray}
These results have been obtained from a purely geometric point of 
view. So far the only physical constraint invoked is 
that the surface stress energy tensor be identically zero. 

When it comes to looking for sources of the Kerr 
metric then the interior metric must correspond to some 
physically sensible
matter distribution. This obviously places further constraints on the
interior metric. One of the most fundamental properties 
we expect from an interior Kerr metric is 
that as $a \rightarrow 0$ the seed metric is recovered and is an interior 
Schwarzschild solution. The recovery of the seed metric for $a=0$ is 
inherent in
the properties of~(\ref{eq:boy}) and the transformations~(\ref{eq:trans}).
However what is not necessarily true, and has not even been
discussed until now, is that the metric~(\ref{eq:boy}) corresponds 
to a physical reasonable 
solution to Einstein's field equations. 

Before proceeding any further it should be made clear by what is meant by the
term ``physically reasonable''. Their are a number of 
criteria that one could 
use. The ones that we are using, as we consider them to 
be the most fundamental, are the following:
\begin{itemize}
\item The strong and weak energy conditions are obeyed, 
i.e. the density 
$\rho$ is 
always positive and the density is always greater than the  
pressure $P$: $\rho \geq 0$; $\rho \geq p$.
\item $P$ and $\rho$ are monotonically decreasing as we move out from the
center.
\item $P$ and $\rho$ are related by a sensible equation of state. 
\item The interior is matched smoothly to the exterior. 
\end{itemize}

With this in mind, we need to provide a seed metric for the 
NJA which satisfies the boundary conditions along with 
the above physical constraints.  
The technique frequently used for  
obtaining  such
interior seed solutions is to combine  
Einstein's field equations 
along with the conservation laws for a given 
stellar model, such as a  
perfect fluid with a given equation of state. In the case of SSS
perfect fluids this
lends to the Oppenheimer-Volkov 
equation~\cite{pap:jro}. Even with these simplifications 
analytic solutions to the Oppenheimer-Volkov equation are
difficult to obtain, see~\cite{bk:dk} for 
some examples. 
It is noted in~\cite{bk:dk} that the few known analytic 
solutions generally fail on physical grounds.
This is a fairly major hurdle in the application of NJA to 
interior solutions since exact analytic expressions are 
required in order to be successfully. 

\section{A trial solution}
\label{sec:ats}              
One of the great difficulties we face in looking for 
solutions sources of the Kerr metric is the matching of 
boundary conditions on 
appropriate surfaces. The approach we are
going to take is not the usual one of guessing the solution 
and then seeing if the boundary conditions match. In fact 
we are going to take the
reverse approach. We will start with interior structures that   
join smoothly to the Kerr metric.
We will then examine the physical properties of these structures 
via the generation of the Einstein tensor. If these structures are 
``physically reasonable'' we will then claim to have found an 
interior solution that may be considered as a source 
for the Kerr metric. 

Although the boundary conditions ~(\ref{eq:k1}) to~(\ref{eq:k4}) 
come in a relatively simple form, they are still 
rather difficult to work with. 
However this situation is rectifiable by considering surfaces 
described by $\partial{R(\theta)}/\partial{\theta}=0$. 
The justification for 
choosing such simplified boundary surfaces goes beyond 
just making the 
equations more easily solvable. To understand this 
the first thing we 
must realise is that the Kerr metric written in Boyer-Lindquist 
coordinates leads to a confusing interpretation of the variable $r$. 
We wish to emphasise the point again here that the four degrees of 
gauge freedom in relativity 
invoke an ambiguity of coordinate 
definitions. Consider for example the Kerr metric 
as first written by Kerr in Cartesian 
coordinates;
\begin{eqnarray}
ds^2 & = & d{\bar t}^2 -dx^2-dy^2-dz^2  \nonumber \\
& - & \frac{2m\varrho^3}{\varrho^4 + a^2z^2}\left[ \frac{\varrho(xdx+ydy)-a(xdy-ydx)}{\varrho^2+a^2}
+ \frac z\varrho dz + d{\bar t}\right]^2,
\end{eqnarray}
where $\varrho$ is determined implicitly, up to a sign, by 
\begin{equation}
\varrho^4 -(x^2 + y^2 + z^2 -a^2)\varrho^2 -a^2z^2 = 0.
\end{equation}
The coordinates $x,y,z$ in the Cartesian form are related to $r,\theta,\phi$ 
in the Boyer-Lindquist form in the following way 
\begin{equation}
x = r\sin{\theta}\cos{\phi} + a\sin{\theta}\sin{\phi};\;
y = r\sin{\theta}\sin{\phi} - a\sin{\theta}\cos{\phi};\;
z = r\cos{\theta}. \label{eq:xyz}
\end{equation}
If we take the magnitude of the radial vector in spherical 
coordinates $(radius)$ to 
have its usual definition,
\begin{equation}
radius^2 = x^2 + y^2 + z^2, \label{eq:r2}
\end{equation}
then by substitution of~(\ref{eq:xyz})
into~(\ref{eq:r2}) we obtain that 
\begin{equation}
radius^2 = r^2 + a^2\sin^2\theta.
\end{equation}
Since a surface is described by $r=R(\theta)$  
 it is apparent why 
$\partial{R(\theta)}/\partial{\theta}=0$, i.e. 
$R(\theta)=R=$ constant, is a sensible choice
of  boundary  surface. The surface defined by 
\begin{equation}\label{eq:ob}
radius^2 - a^2\sin^2\theta = R^2
\end{equation}
is and oblate spheroid. 
The oblate spheroid is a surface of revolution 
swept out by an ellipse 
rotating about its minor axis. An oblate spheroid
defined by eq.(\ref{eq:ob}) 
transforms to a sphere in the limit $a\rightarrow0$.
Clearly these are the types of surfaces one would expect 
for a uniformly rotating star.  
If, as we have argued, the boundary surface separating the 
interior and exterior solutions is an oblate spheroid then 
equations~(\ref{eq:k1}) 
to~(\ref{eq:k5}) simplify to 
\begin{eqnarray}
e^{2\phi(r,\theta)-}\!|_s & = & 1 - \frac{2MR}{\Sigma_s}
 \label{eq:ep}\\
e^{2\lambda(r,\theta)-}\!|_s & = & \frac{\Sigma_s}{\Sigma_s - 
2MR}  \label{eq:el} \\
\frac{2M(R^2-a^2\cos^2{\theta})}{\Sigma^2_s} & = &
e^{2\phi(r,\theta)-},_r\!|_s  \label{eq:epr} \\
\frac{2M(a^2\cos^2\theta-R^2)}{\Delta^2_s} & = & 
e^{2\lambda(r,\theta)-},_r\!|_s.  \label{eq:elr} 
\end{eqnarray}
The above equations give the boundary conditions for the joining of 
a SAS metric generated by the NJA to the 
Kerr metric on a static oblate spheroidal boundary surface. 
With these 
constraint equations it possible to look for new sources of the Kerr metric. 

The first solution of this kind we examine, and the one which is perhaps 
the simplest, is
\begin{equation}
e^{2\lambda(r,\theta)-}= \frac{\Sigma^2_s -2M(R^2r + 
a^2\cos{\theta}(2R-r))}{\Delta^2_s}. \label{eq:e2lam}
\end{equation}
Using this guess it should be possible to obtain, a density and 
pressure profile as well as finding an exact solution for 
$e^{2\phi(r,\theta)}$ by the use of Einstein's equations. 
It is shown by the extremely low number of exact solutions 
Einstein's equations
that such a method, even for SSS metrics, is
extremely difficult to work with. The use of algebraic programmes 
such as 
Cartan and grtensor~\cite{prog:pm}~\cite{prog:hhs} 
cut down most of the algebra however they 
do not reduce the problem to a solvable state. 
%As an example of the 
%the difficulty in solving Einstein's field equations the interested 
%reader is referred Chandrasekhar's book~\cite{bk:sc}
%in which the Kerr solution is 
%obtained with out assuming that the metric has any 
%particular form.

Although it is a complex task to examine the nature of stationary 
axially symmetric metrics the examination of the non-rotating case is
much simpler. It is always useful to examine the slow or 
zero rotation limit of such solutions since one would expect that 
these limits must also represent physical objects. 
By examining the $a\rightarrow0$ limit
of such metrics we are examine the properties of the seed metric. Recall that 
in the previous section we showed that the NJA could not be 
successfully applied to  
{it any} arbitrary seed metric. 

We know that our trial solution for $e^{2\lambda(r,\theta)}$ given 
by~(\ref{eq:e2lam})
obeys the boundary conditions on an oblate spheroid surface. 
One method for obtaining the interior structure might be to examine
various functions of $e^{2\phi(r,\theta)}$ that obey the boundary
conditions~(\ref{eq:ep}) and~(\ref{eq:epr}). Once this has been done 
these functions along with~(\ref{eq:e2lam}) could be fed into the 
general metric~(\ref{eq:boy}), from this the stress-energy tensor may be
obtained by equating it to the Einstein tensor. However the authors are 
opposed to such a method as it relies on a great deal of guess work. 
As such it does not guarantee anything about the physical nature of the 
objects we are examining. 

A more sensible approach, we believe, is to examine the properties of the
seed metric first. If the seed metric corresponds to a physically sensible 
object then the hope is that the new metric generated by the NJA 
will also represent a physical object. 
To begin this process the $a \rightarrow 0$ limit is taken of 
the metric~(\ref{eq:boy}) so that is~(\ref{eqn:ssm})
and~(\ref{eq:e2lam}) becomes
\begin{equation}
e^{2\lambda(r_*)} = \frac{1-xr_*}{(1-x)^2}, \label{eq:e2lama}
\end{equation}
where $x = 2M/R$ and 
$r_* = r/R$.

The theory of radially symmetric distributions of matter is 
a well explored 
field. One of the classic papers on this subject is by
Wyman~\cite{pap:mw}. For an isotropic fluid sphere described
by~(\ref{eqn:ssm}) the pressure 
$P$ and density $\rho$ satisfy the relations
\begin{eqnarray}
P_* & = & e^{-2\lambda}(2\phi'/r_* +1/r_*^2)-1/r_*^2\label{eq:P1} \\ 
P_* & = & e^{-2\lambda}(2\phi''/r_* -\lambda'\phi'
+(\phi'-\lambda')/r_* + \phi'^2) \label{eq:P2} \\
\rho_* & = & e^{-2\lambda}(2\lambda'/r_* -1/r_*^2)+1/r_*^2 
\label{eq:dens} \\
P'_* & = & -(P_*+\rho_*)\phi' \label{eq:emq}
\end{eqnarray} 
The notation used is that $\lambda$ and $\phi$ are understood to 
be only functions of $r_*$, $P_*=PR^2$, $\rho_*=\rho R^2$ and
the prime denotes derivatives with respect to $r_*$. 
The equality of equations~(\ref{eq:P1}) and~(\ref{eq:P2}) 
imply~(\ref{eq:emq}). From~(\ref{eq:dens}) 
and~(\ref{eq:e2lama}) the density 
profile is 
\begin{equation}
\rho_* = \frac1{r_*^2} - \frac{(1-x)^2}
{r_*^2(r_*x-1)^2}.
\end{equation}
This density profile corresponds to that of a 
physically sensible stellar object as 
it as always 
positive and monotonically decreasing 
for all values of $x$ such that 
$0<x<1$. For this model the density vanishes at the surface 
for all allowed
values of $x$. 

Given that the function $e^{-2\lambda}$ is known then the 
equality of~(\ref{eq:P1}) and~(\ref{eq:P2}) results in a 
Riccati equation 
of first order in $\phi'$. There are various methods for finding the 
solutions such equations~\cite{bk:gmm} 
and once this is done the pressure profile may be
obtained. For the example given the resulting Riccati equation is
\begin{equation}
\phi'' = \frac1{r_*^2} + \frac{r_*x-1}{r_*(x-1)^2} + 
\frac{x}{2r_*(1-r_*x)} +
\left[ \frac1{r_*} + \frac{x}{2(r_*x-1)} \right]\phi' 
- \phi'^2.
\end{equation}
Since this is a second order differential equation the 
solution involves two 
constants of integration. These two constants are determined by 
making sure that at the surface 
the pressure is zero  and that  the
metric is Schwarzschild. In general the solutions 
are defined by
functions more complicated than the elementary 
transcendental functions. 
Thus far we have tried without success to obtain analytic 
solutions. As a result we have decided to perform numerical 
integration to 
examine the solutions quantitatively. 

The integration routine we have chosen to use is the ubiquitous 
fourth-order
adaptive step-size Runge-Kutta routine. Using this routine we 
are able to 
see how $\phi'$ varies as a function of $r_*$. This along with the 
the know relation $e^{-2\lambda}$ enables us to 
determine the pressure 
profile of the stellar object. The pressure and density profiles 
are shown 
in figures~(\ref{fig:dens}) to~(\ref{fig:pressvdens}) 
for $x=0.3$ as typical for neutron stars.   

\section{conclusion}
\label{sec:c}	
To make it clear what new discoveries and investigations 
have been made in 
this paper it is important to make a brief summary. Recall 
that in the first 
section~\ref{sec:i} we gave a history to the discovery of the 
Kerr metric and of the
Newman-Janis algorithm (NJA). We discussed the fact that the 
formalism has 
been well developed in the context of metrics that can be written in 
Kerr-Schild (KS) form. Furthermore we pointed out that 
interior solutions, 
with the exception of the pure radiation solutions, 
can not been written in the KS form. 
Essential this means that the NJA as applied to interior solutions has 
not been appropriately explored. With this in mind we set about 
the task 
of rectifying this situation. 
The second section~\ref{sec:tnja}  commenced 
by generalising the Newman-Janis algorithm for any 
static spherically symmetric (SSS) ``seed'' metric. We 
pointed out that 
although this has previously been investigated by others 
its lack of
success may be attributed to an inconvenient choice of coordinates 
and to
boundary conditions which are too stringent. To eliminate this 
problem  
a coordinate transformation was made after applying the NJA 
so that this new 
metric could be written in Boyer-Lindquist type form. This result to
the best of the authors knowledge  
is completely new. The metric formed by this algorithm belongs to 
 a special class of metrics 
which are both stationary and axially symmetric (SAS). 
The Kerr metric 
belongs to this class. 

In the third section~\ref{sec:tmotstoacbs} we 
established the constraints placed 
by matching smoothly two such metrics on any static axially 
symmetric surface. 
The boundary conditions we used were those first developed Dormois 
and Israel. We found that for metrics generated by the NJA 
the boundaries 
conditions are a relatively simple set of constraint equations. 
  
In looking for interior solutions to rotating massive bodies one 
would expect 
that the exterior to be described by the Kerr metric. 
Section~\ref{sec:psftkm} 
therefore was devoted to examining the properties of 
an interior  
metric that matches smoothly to the Kerr metric. Also in 
this section 
we elaborated on what is meant by the term 
``physically reasonably'' in 
describing the properties of stellar objects. 

Finally in the last section before the conclusion~\ref{sec:ats}
we determined what 
sort of interior metrics could be matched smoothly to the 
Kerr metric 
on oblate spheroidal surfaces. We found one such solution we have termed
the trial solution in which the metric coefficient $e^{\lambda(r,\theta)}$
was determined explicitly. We argued that although it would be just as
simple to 
find functions of $e^{\phi(r,\theta)}$ that matched the boundary 
conditions on such surfaces, one has no a priori reason to believe that 
these will correspond to a physically reasonable stellar model. 
To work around this 
problem we decided that to begin with a physically reasonable seed
metric. To make sure this was the case we took the zero 
rotation limit $(a=0)$ of $e^{\lambda(r,\theta)}$. We found that the 
density profile resulting from this was physical 
sensible. The density $\rho$ as a function of the radius $r$ was 
always positive,  decreased  monotonically
out from the center and although it was infinite at the center
the total mass was still finite. 

The pressure profile on the other hand  was a slightly 
more difficult to explicate. We started by looking at seed 
metrics which 
described perfect fluids with isotropic pressure. Although 
this simplified 
calculations greatly it meant that in order to find exact solutions
of $e^{\phi}$ we needed to integrate a first order non-liner 
equation known as the Riccati equation. The task of finding such 
solutions is non trivial as solutions generally 
can not expressed as 
simple transcendental functions. However such equations can 
be integrated
numerically without too much effort. 

Once this was done it was possible to determine the pressure 
profiles for 
various mass to surface radius ratios ($x\equiv2M/R$). 
An examination of 
one such profile for a typical value $x$ for compact objects namely
$x=0.3$ showed extremely encouraging results. The pressure 
$P_*$ profile 
as a function of normalised radius $r_*$ satisfied all 
our specified criteria for 
being physically reasonable. The pressure was a 
monotonically decreasing, the strong 
energy condition was obeyed and although
the pressure diverged at $r=0$ the amount of energy within a 
given volume was still finite. 

>From the original investigations made in this 
paper we can confidently state that 
the NJA may be applied to find new sources of the 
Kerr metric. We have shown that there exists  
physically sensible 
seed metrics to which when the NJA is applied new SAS
metrics are generated. These metrics are  
considered as sources 
of the Kerr metric. 
The new 
solutions have continuous boundary conditions on  
oblate spheroidal surfaces.
At present the physically 
properties of these new metrics has not been fully investigated. 
The sole reason for this being 
that it is a somewhat arduous task. However we are currently 
embarking on this project.
The authors feel confident that further work in this 
area will complete the task of obtaining 
the long sort after sources for the Kerr metric. 

\section{acknowledgements}
SPD would like to thank the Australian Postgraduate Programme 
for support during the completion of this work. SPD would 
also like to thank the University of Padova for its hospitality
during his stay there in which most of this work was done.  
SPD and RT would also like to thank (in alphabetical order) 
N. J. Cornish, F. De Felice, N. E. Frankel,
G. Magli and P. Szekeres for many helpful discussion. We would 
also like to thank C. P. Dettmann for providing the skeleton 
programme for the fourth-order Runge-Kutta routine.

\begin{figure}
\caption{The density profile of the trial solution with 
$a=0$ and $x=0.3$}
\label{fig:dens}
\end{figure}

\begin{figure}
\caption{The pressure profile of the trial solution with 
$a=0$ and $x=0.3$}
\label{fig:press}
\end{figure}

\begin{figure}
\caption{Combined density and pressure profiles of the trial solution with 
$a=0$ and $x=0.3$}
\label{fig:pressanddens}
\end{figure}

\begin{figure}
\caption{Pressure versus density plot for the trial solution with 
$a=0$ and $x=0.3$}
\label{fig:pressvdens}
\end{figure}

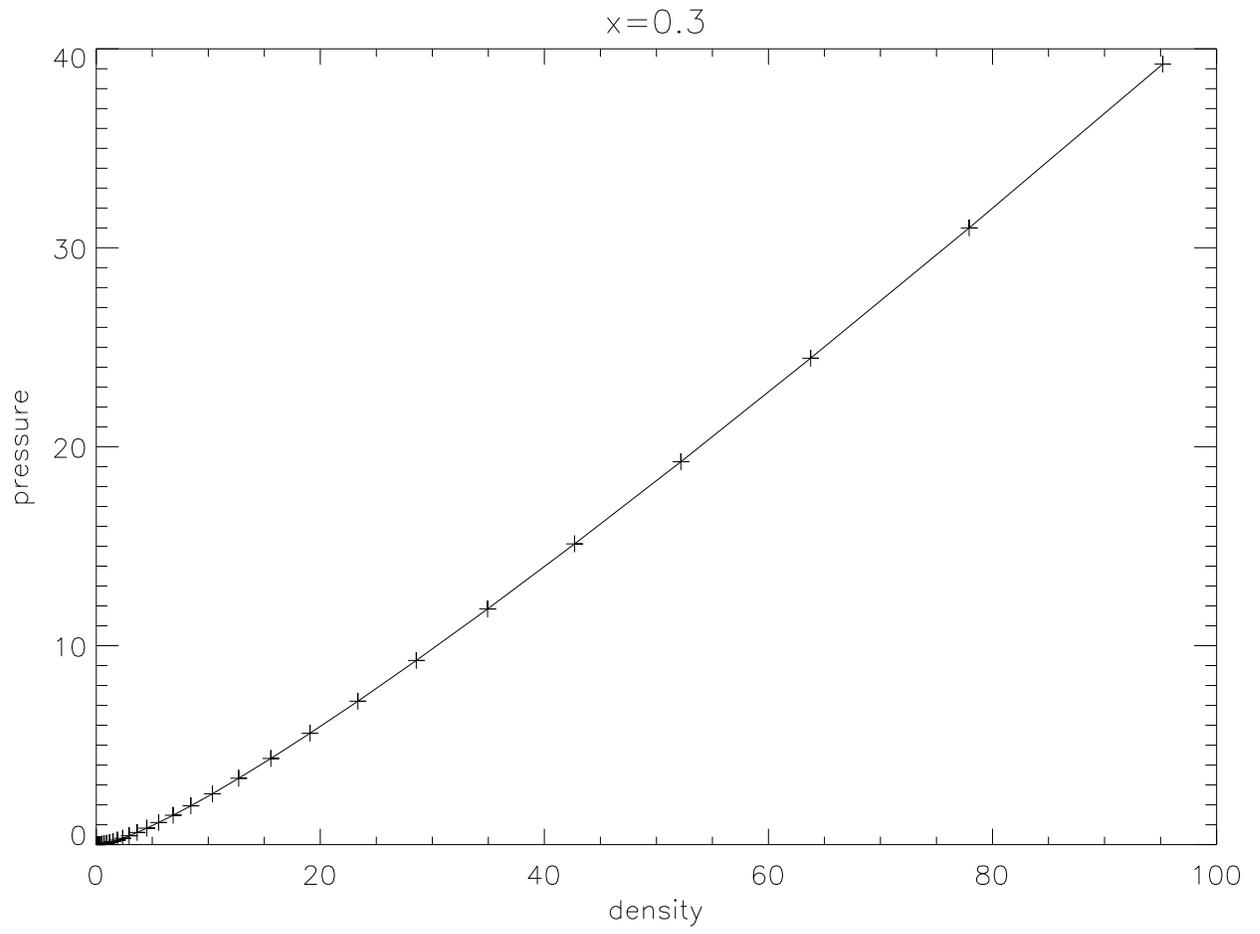
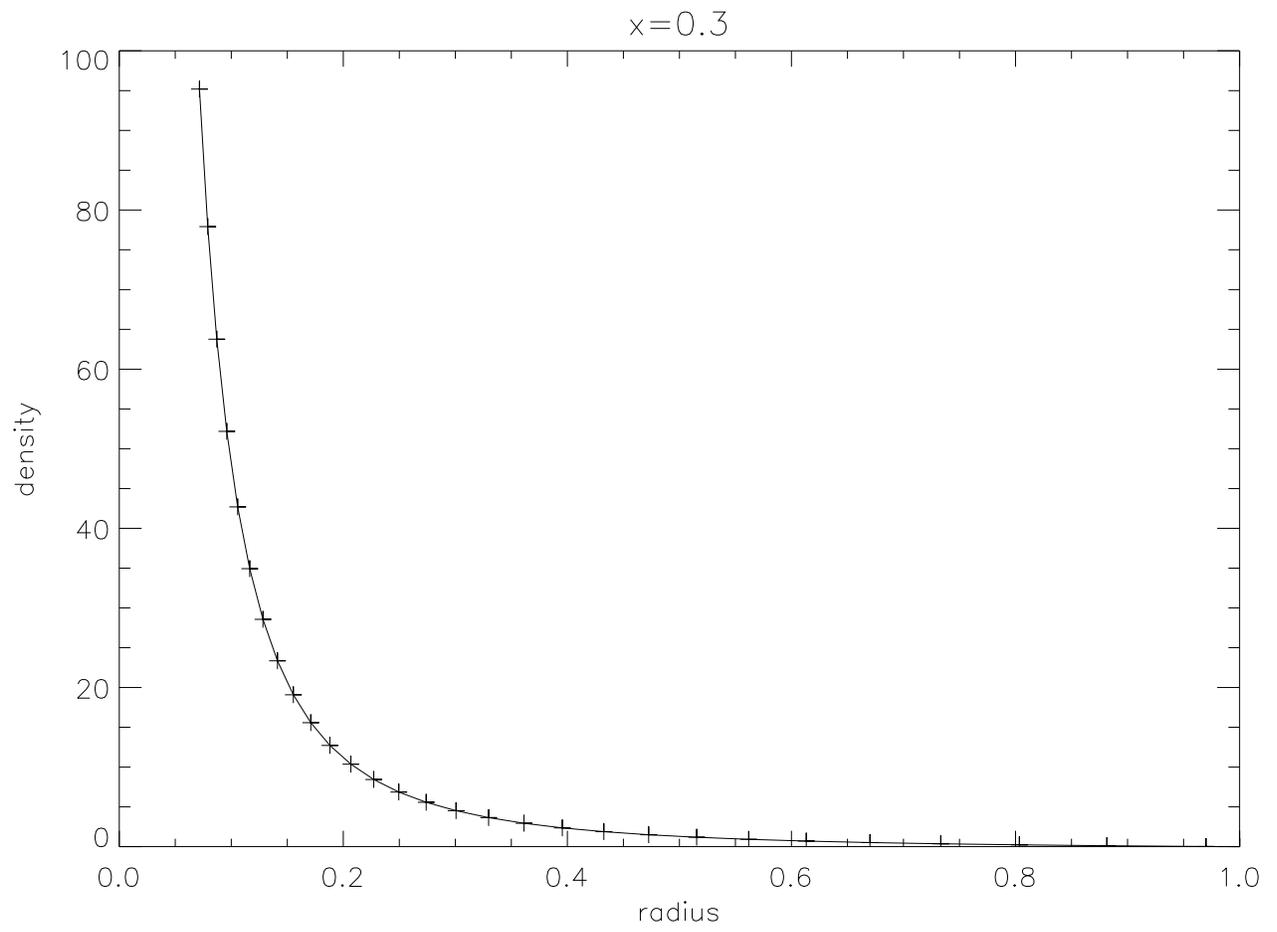
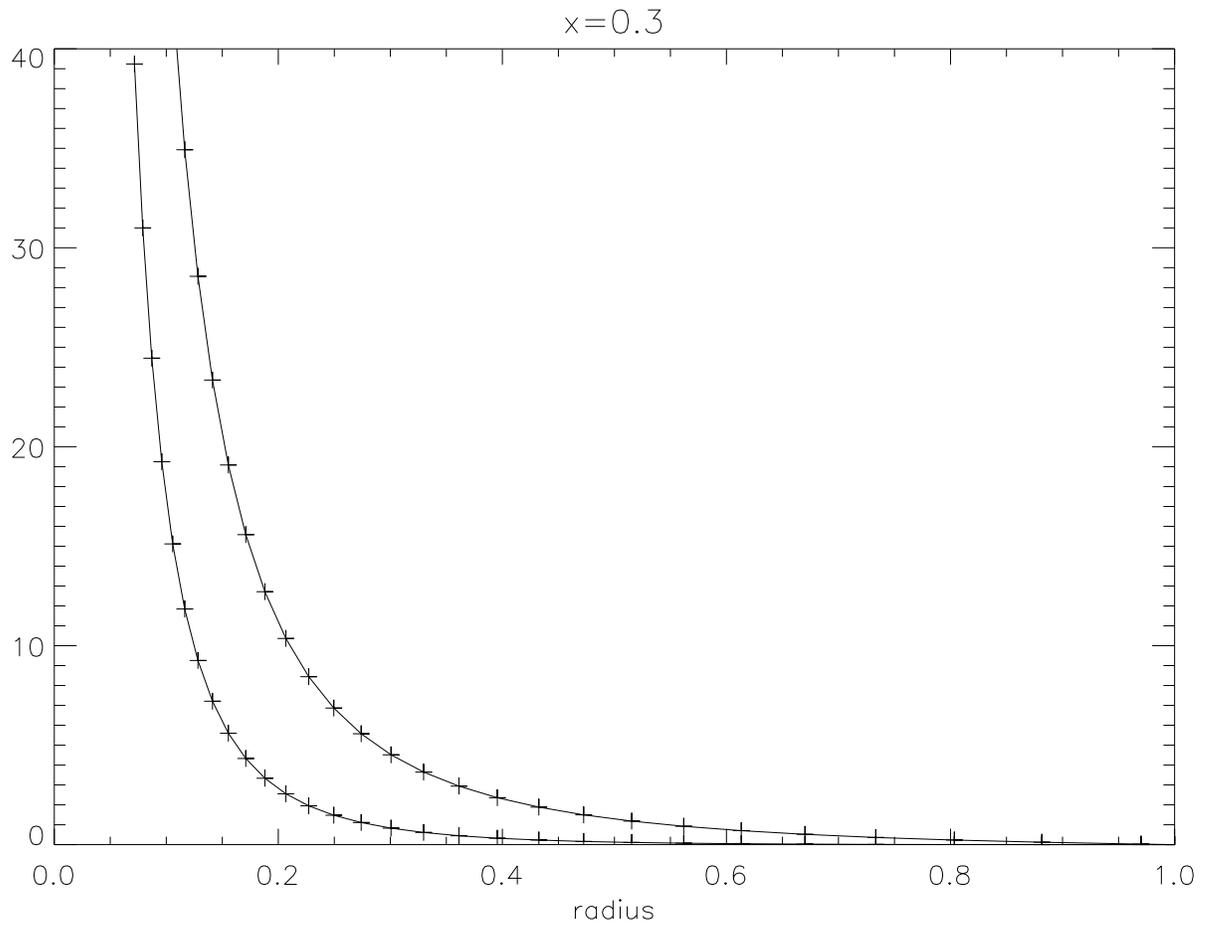
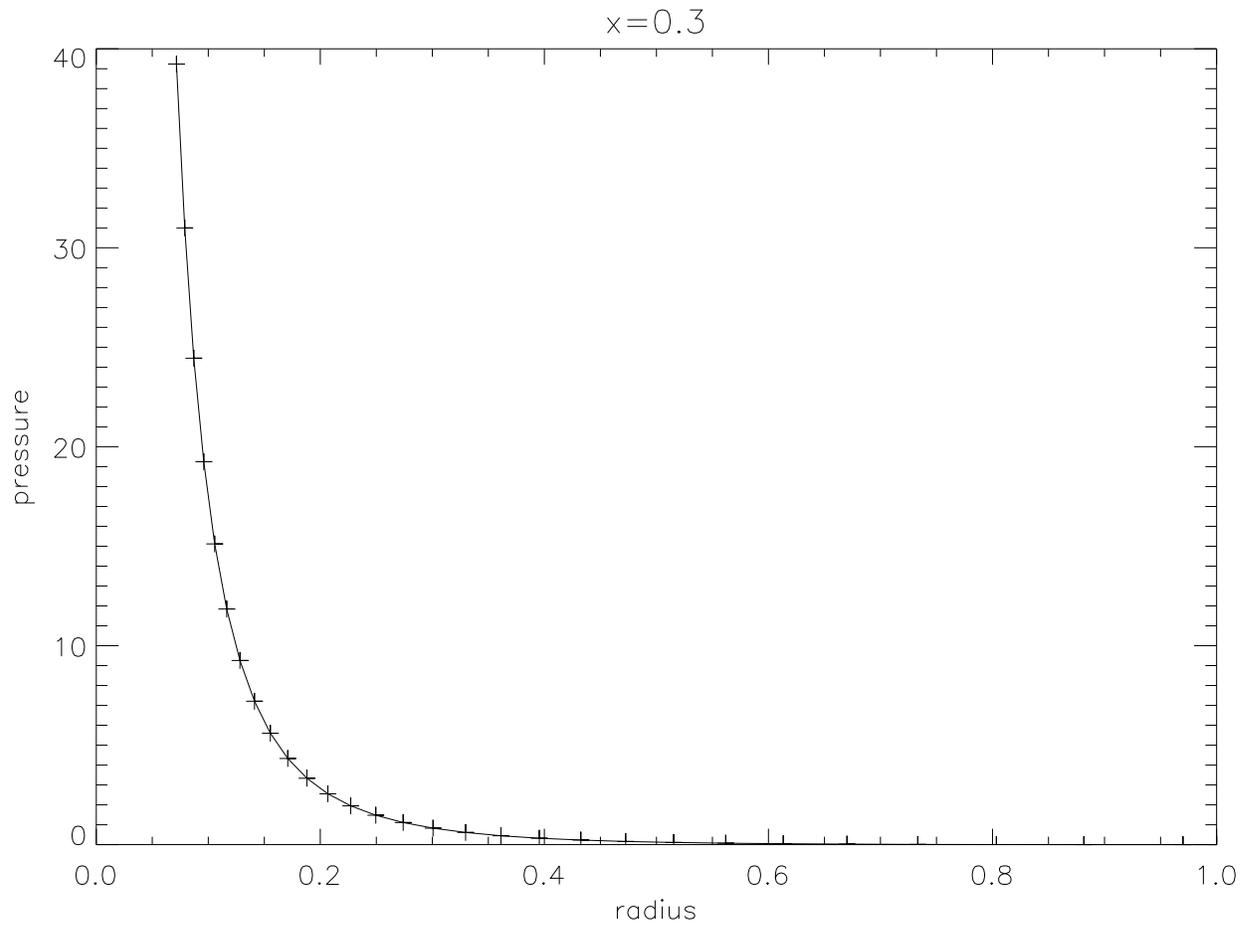
\end{document}